\documentclass[twoside,twocolumn,english,pre]{revtex4}
\usepackage[T1]{fontenc}
\usepackage[latin1]{inputenc}
\usepackage{amsmath}
\usepackage{graphicx}
\usepackage{amssymb}

\makeatletter

\providecommand{\LyX}{L\kern-.1667em\lower.25em\hbox{Y}\kern-.125emX\@}

\usepackage{babel}
\makeatother
\begin{document}

\title{Energy Spectrum Evolution of a Diffuse Field in Elastic Body Caused
by Weak Nonlinearity}

\author{Alexei Akolzin}

\email{akolzine@uiuc.edu}

\author{Richard L. Weaver}

\email{r-weaver@uiuc.edu}

\affiliation{Department of Theoretical and Applied Mechanics, University of Illinois,
104 S. Wright Street, Urbana, Illinois 61801, USA}

\date{\today}

\begin{abstract}
We study the evolution of diffuse elastodynamic spectral energy density
under the influence of weak nonlinearity. It is shown that the rate
of change of this quantity is given by a convolution of the linear
energy at two frequencies. Quantitative estimates are given for sample
aluminum and fused silica blocks of experimental interest.
\end{abstract}

\pacs{05.45.Mt, 43.25.+y, 43.35.+d, 62.30.+d}

\maketitle

\section{Introduction}

Weak nonlinearity is known to cause a redistribution of elastodynamic
spectral energy density. Energy present at one or two base frequencies
will migrate, under the influence of nonlinearity, to higher harmonics
and to sum and difference frequencies. If those sum and difference
frequencies originally have little or no energy, their presence can
be a sensitive measure of nonlinearity. The method has been shown
to be capable of detecting flaws in otherwise linear-response specimens.
It has been applied to crack detection \cite{ref:Sutin-Zaitsev} and
to assessing the quality of adhesive bonds \cite{ref:Wegner-Rothenfusser}.
Much of this work has been qualitative, with emphases on how the nonlinear
effects scale with field amplitude. The geometry is such, and the
fields sufficiently uncontrolled, that attempts to quantify the nonlinearity
in absolute terms have not been feasible.

In special circumstances, with high amplitude and long distance plane
wave propagation, it is possible to monitor harmonic generation, and
thereby assess nonlinearity, and do so quantitatively. It has been
suggested that nonlinear Rayleigh wave propagation may be a useful
NDE technique to measure surface properties \cite{ref:Hurley}. Lamb
waves were also suggested for this purpose \cite{ref:deLima}.

As an alternative to plane wave configurations diffuse elastodynamic
fields might be appropriate for NDE measurements under insufficiently
controlled conditions. Diffuse fields span a broad range of applications
from optics and microwaves to ultrasonics (see \cite{ref:Stockmann}
and references therein). Gross properties of a specimen can be evaluated
from integral field parameters \cite{ref:Pine-Weaver}. Extension
of diffuse field theory to include nonlinear effects is receiving
increasing experimental \cite{ref:Boer} and theoretical \cite{ref:Bressoux-Skipetrov}
attention. Its application to evaluation of material properties may
be a future research focus.

The purpose of this paper is to describe effects of weak nonlinearity
on the energy spectrum of a diffuse field, as one of the integral
quantities available for experimental measurement. With a view toward
eventual application in the context of NDE, but with an interest in
the general problem, we take elastodynamics in a nonlinear solid as
an underlying physical system for our diffuse field model. 

We start with the governing equations for the chosen field type in
Section II, and develop them into a system of ODEs with quadratic
and cubic nonlinear terms describing time-evolution of the modal amplitudes.
In Section III we discuss statistical assumptions for the normal frequencies
and modes of the solid. In Section IV the nonlinear equations are
treated by means of regular perturbation theory and averaged to obtain
the spectral power transfered into a frequency band. Results independent
of the underlying physical nature of the nonlinearity are considered.
In Section V we specialize to the case of an isotropic homogeneous
elastic body and an initial field generated by two transient narrow-band
signals centered at different frequencies. Quantitative estimates
for the nonlinear energy transfer into double and combination frequencies
are provided and discussed. Conclusions are presented in Section VI.

\section{Governing equations}

Suppose we have a finite hyperelastic body occupying volume $V_{0}$
in its natural state referenced by coordinate $\mathbf{x}$, and having
material density $\rho _{0}\left(\mathbf{x}\right)$. Strain energy
density of the body for a displacement field $\mathbf{u}$ with a
corresponding Green's tensor $E_{ij}=1/2\left(u_{i,j}+u_{j,i}+u_{k,i}u_{k,j}\right)$
is described to leading orders in strain by \cite{ref:OgdenBook}:
\begin{align}
W= & \frac{1}{2!}C_{ijkl}\left(\mathbf{x}\right)E_{ij}E_{kl}\nonumber \\
 & +\frac{1}{3!}D_{ijklmn}\left(\mathbf{x}\right)E_{ij}E_{kl}E_{mn}\nonumber \\
 & +\frac{1}{4!}F_{ijklmnpq}\left(\mathbf{x}\right)E_{ij}E_{kl}E_{mn}E_{pq}+\ldots \; ,\label{eq:EnergyDensity}
\end{align}
 where $\mathbf{C}$, $\mathbf{D}$ and $\mathbf{F}$ are linear,
second and third-order nonlinear elastic tensors respectively.

Let the body be subjected to external forces and tractions that stop
acting after a certain cutoff time $t_{0}=0$ after which the field
in the body freely evolves without dissipation under zero-displacement
boundary conditions. According to the Hamilton's Principle, the weak
form of the governing equation for the field evolution can be written
as

\begin{equation}
\iint _{V_{0}}\left(-\sigma _{ji}\delta u_{i,j}-\rho _{0}\ddot{u}_{i}\delta u_{i}\right)\, d\mathbf{x}^{3}dt=0,\label{eq:Governing}\end{equation}
 where $\sigma $ is the first Piola-Kirchoff stress tensor. It is
related to the strain energy density (\ref{eq:EnergyDensity}) via
deformation gradient $F_{ij}=\delta _{ij}+u_{i,j}$ \cite{ref:OgdenBook}:\begin{equation}
\sigma _{ji}\left(\mathbf{x},\mathbf{u}\right)=\partial W/\partial F_{ij}.\label{eq:PiolaTensor}\end{equation}

To distinguish parts of equation (\ref{eq:Governing}) responsible
for linear and leading-order nonlinear behavior of the system we expand
the stress tensor $\sigma $ in powers of $\left\Vert \mathbf{u}\right\Vert $
up to the third-order, and label the respective linear and nonlinear
operators as $\widehat{\mathbf{L}}$ and $\widehat{\mathbf{N}}$.
Equation (\ref{eq:Governing}) is then written as \begin{equation}
\iint _{V_{0}}\left[\delta u_{i}\left(\rho _{0}\ddot{u}_{i}-\widehat{L}_{i}\mathbf{u}\right)+\delta u_{i,j}\widehat{N}_{ij}\mathbf{u}\right]d\mathbf{x}^{3}dt=0.\label{eq:WeakForm}\end{equation}

A complete set of eigenvalues $\omega _{n}^{2}$ and their normalized
eigenfunctions $\mathbf{u}^{n}$ is considered to be known for the
linear operator as solution of \begin{equation}
\widehat{L}_{i}\mathbf{u}^{n}\equiv \frac{\partial }{\partial x_{j}}C_{ijkl}u_{k,l}^{n}=-\rho _{0}\omega _{n}^{2}u_{i}^{n}.\label{eq:EigenProblem}\end{equation}

For clarity of notation we introduce composite Greek indices with
implied summation upon them:\begin{eqnarray*}
g_{\alpha \equiv \left\{ \mathbf{x},i,j\right\} }h_{\alpha \equiv \left\{ \mathbf{x},i,j\right\} } & = & \int \limits _{V_{0}}g_{ij}\left(\mathbf{x}\right)h_{ij}\left(\mathbf{x}\right)d\mathbf{x}^{3},
\end{eqnarray*}
and denote the first partial derivatives of the field as separate
functions: $u_{\alpha \equiv \left\{ \mathbf{x},i,j\right\} }=\partial u_{i}\left(\mathbf{x}\right)/\partial x_{j}$.

The displacement field allows decomposition upon the eigenfunctions
with time-dependent modal amplitudes $d_{k}$ \cite{foot:BC}:\[
u_{i}\left(\mathbf{x},t\right)=\sum _{k}d_{k}\left(t\right)u_{i}^{k}\left(\mathbf{x}\right).\]
By employing this representation in equation (\ref{eq:WeakForm}),
and using the eigenfunctions as a set of test functions $\delta \mathbf{u}$,
we restate the governing equation as \begin{align}
\ddot{d_{k}}+\omega _{k}^{2}d_{k}= & -\sum _{m,l}N_{\alpha \beta \gamma }U_{\alpha \beta \gamma }^{klm}d_{k}d_{l}d_{m}\nonumber \\
 & -\sum _{m,l,n}N_{\alpha \beta \gamma \delta }U_{\alpha \beta \gamma \delta }^{klmn}d_{k}d_{l}d_{m}d_{n},\label{eq:ODE}
\end{align}
where $U_{\alpha \beta \gamma }^{klm}=u_{\alpha }^{k}u_{\beta }^{l}u_{\gamma }^{m}$,
and $U_{\alpha \beta \gamma \delta }^{klmn}=u_{\alpha }^{k}u_{\beta }^{l}u_{\gamma }^{m}u_{\delta }^{n}$.
All the specifics of particular type of nonlinear behavior are now
contained in modal matrices $\mathbf{NU}$.

Only symmetrical with respect to the last two indices parts of matrix
$N_{\alpha \beta \gamma }$ and symmetrical with respect to the last
three indices parts of matrix $N_{\alpha \beta \gamma \delta }$ survive
modal summation in (\ref{eq:ODE}). Without loss of generality we
put these matrices equal to their corresponding symmetrical parts,
and write them in Greek index notation as \begin{align*}
N & _{\alpha \equiv \left\{ \mathbf{x},i,j\right\} \beta \equiv \left\{ \mathbf{x}',k,l\right\} \gamma \equiv \left\{ \mathbf{x}'',m,n\right\} }\\
 & =\frac{1}{2}N_{ijklmn}\delta \left(\mathbf{x}-\mathbf{x}'\right)\delta \left(\mathbf{x}-\mathbf{x}''\right),
\end{align*}
 and \begin{align*}
N & _{\alpha \equiv \left\{ \mathbf{x},i,j\right\} \beta \equiv \left\{ \mathbf{x}',k,l\right\} \gamma \equiv \left\{ \mathbf{x}'',m,n\right\} \delta \equiv \left\{ \mathbf{x}''',p,q\right\} }\\
 & =\frac{1}{6}N_{ijklmnpq}\delta \left(\mathbf{x}-\mathbf{x}'\right)\delta \left(\mathbf{x}-\mathbf{x}''\right)\delta \left(\mathbf{x}-\mathbf{x}'''\right).
\end{align*}
 Nonlinear terms of the stress (\ref{eq:PiolaTensor}) yield directional
tensors in the above formulae:\begin{align}
N & _{ijklmn}\nonumber \\
 & =C_{ijln}\delta _{km}+C_{jnkl}\delta _{im}+C_{jlmn}\delta _{ik}+D_{ijklmn},\label{eq:NTensor}
\end{align}
 and \begin{align*}
N & _{ijklmnpq}\\
 & =C_{jlnq}\delta _{ik}\delta _{mp}+C_{jnlq}\delta _{im}\delta _{kp}+C_{jqnl}\delta _{ip}\delta _{mk}\\
 & +D_{jlmnpq}\delta _{ik}+D_{jnklpq}\delta _{im}+D_{jqmnkl}\delta _{ip}\\
 & +D_{ijlnpq}\delta _{mk}+D_{ijlqmn}\delta _{pk}+D_{ijqnkl}\delta _{pm}\\
 & +F_{ijklmnpq}.
\end{align*}
 Due to the major and minor symmetries of elastic tensors \textbf{$\mathbf{C}$},
$\mathbf{D}$ and $\mathbf{F}$ \cite{ref:OgdenBook} both matrices
$\mathbf{N}$ can be identified as fully symmetric:\begin{eqnarray}
 & N_{\alpha \beta \gamma }=N_{\beta \alpha \gamma }=N_{\gamma \beta \alpha }, & \nonumber \\
 &  & \label{eq:NSymmetry}\\
 & N_{\alpha \beta \gamma \delta }=N_{\beta \alpha \gamma \delta }=N_{\gamma \beta \alpha \delta }=N_{\delta \beta \gamma \alpha }. & \nonumber 
\end{eqnarray}

A quantity which we term linear energy stored in a single mode $k$
at time $t$ is \begin{equation}
E_{k}=\frac{1}{2}\left(\dot{d}_{k}^{2}+\omega _{k}^{2}d_{k}^{2}\right).\label{eq:E}\end{equation}
In the absence of nonlinearity it is equal to the total energy of
a mode, and is constant over time. It can be written in terms of the
complex amplitudes $\psi _{k}$:\begin{equation}
E_{k}^{\left(0\right)}=\frac{1}{2}\left|\psi _{k}\right|^{2}\omega _{k}^{2}.\label{eq:E0}\end{equation}
The amplitudes $\psi _{k}$ arise from the action of external forces
and tractions prior to the cutoff time $t_{0}$. They describe evolution
of the linear part of the field by means of the modal amplitudes $d_{k}^{\left(0\right)}$
that are found as solutions of the linearized version of equation
(\ref{eq:ODE}), with matrices $\mathbf{N}$ put to zero \cite{ref:Weaver84}:\begin{equation}
d_{k}^{\left(0\right)}\equiv a_{k}\left(t\right)=\Im \psi _{k}e^{-\imath \omega _{k}t}.\label{eq:a}\end{equation}

The energy flow due to nonlinearity, $\dot{E}_{k}=\Pi _{k}$, is obtained
from the governing equation (\ref{eq:ODE}), with modal power input
being\begin{align}
\Pi _{k}= & -N_{\alpha \beta \gamma }\sum _{m,l}U_{\alpha \beta \gamma }^{klm}\dot{d}_{k}d_{l}d_{m}\nonumber \\
 & -N_{\alpha \beta \gamma \delta }\sum _{m,l,n}U_{\alpha \beta \gamma \delta }^{klmn}\dot{d}_{k}d_{l}d_{m}d_{n}.\label{eq:PiModal}
\end{align}
 It must be noted that the energy quantity $E_{k}$ (\ref{eq:E})
is not conserved, because it is not a true energy in the nonlinear
case. Analysis of the strain energy density (\ref{eq:EnergyDensity})
reveals that additional terms $-N_{\alpha \beta \gamma }\sum _{m,l}U_{\alpha \beta \gamma }^{klm}d_{k}d_{l}d_{m}/3-N_{\alpha \beta \gamma \delta }\sum _{m,l,n}U_{\alpha \beta \gamma \delta }^{klmn}d_{k}d_{l}d_{m}d_{n}/4$
must be added to $E_{k}$ to produce a quantity that is conserved.
However, the terms involve summation upon modal amplitudes other than
that of the mode at hand, and do not allow simple interpretation in
terms of a single mode. Thus they are not used.

\section{Statistical model}

We restrict ourselves to a class of systems for which the field excited
by external forces and tractions has a fully diffuse nature. In experiments
such fields are practically realized, for example, in an elastic solid
of a classically chaotic shape, and have statistical properties close
to or indistinguishable from those of a field described by a random
Hamiltonian \cite{ref:Weaver89,ref:Ellegaard-Schaadt}. The normal
modes of the system are taken to be centered Gaussian vectors with
a certain spatial correlation, as was first theoretically conjectured
\cite{ref:Berry-OConnor}, and later numerically and experimentally
verified \cite{ref:McDonald-Schaadt}. We assume that the mean density
of states of the normal frequencies is given by the function $D\left(\omega \right)$
in the form of Weyl-series \cite{ref:Weaver89}, and frequency-frequency
correlations can be neglected.

Since experimental identification of a particular mode is complicated
by the ''missing level'' effect or modal overlap we choose to pursue
calculation of average spectral density rather than individual modal
amplitudes. In order not to distinguish between individual modes at
the frequencies of interest, and thus deal with the average quantities,
we limit observation time of the system $t$ to be less than the corresponding
break (Heisenberg) time, $t_{H}=2\pi D$. On the other hand, time
$t$ is considered larger than transition (ballistic) time in the
solid (mean time between two successive scattering events at the boundaries),
$t_{l}=l/c$, so that diffuse regime of the field is established.
$l$ and $c$ stand for characteristic diameter of and wavespeed in
the solid. Putting together the two bounds yields $t_{l}\ll t\ll t_{H}$,
a condition that can be experimentally realized.

In the framework of the adopted statistical model the modes $\mathbf{u}^{k}$
of the linear operator (\ref{eq:EigenProblem}) are centered Gaussian
random vectors with variance given by\[
\left\langle u_{i}^{k}\left(\mathbf{x}\right)u_{j}^{n}\left(\mathbf{x}'\right)\right\rangle =\delta _{kn}K_{ij}\left(\omega _{k},\mathbf{x},\mathbf{x}'\right),\]
where $\left\langle \ldots \right\rangle $ represents ensemble average,
and $\mathbf{K}$ is a smooth frequency-dependent correlation matrix.
Pairwise correlation of the first partial derivatives of the modes\begin{equation}
\left\langle u_{\alpha }^{k}u_{\beta }^{n}\right\rangle =\delta _{kn}K'_{\alpha \beta }\left(\omega _{k}\right)\label{eq:UPrimeStatistics}\end{equation}
 is readily obtained from \begin{equation}
K'_{\alpha \equiv \left\{ \mathbf{x},i,l\right\} \beta \equiv \left\{ \mathbf{x}',j,m\right\} }=\partial ^{2}K_{ij}\left(\omega ,\mathbf{x},\mathbf{x}'\right)\bigm /\partial x_{l}\partial x'_{m}.\label{eq:KPrimeCalc}\end{equation}
 Orthonormality of the modes imposes a normalization condition upon
the correlation matrix: \begin{equation}
\int _{V_{0}}\rho _{0}\left(\mathbf{x}\right)K_{ii}\left(\omega ,\mathbf{x},\mathbf{x}\right)d\mathbf{x}^{3}=1.\label{eq:KNorm}\end{equation}

From the mode statistics the complex amplitudes $\psi _{k}$ describing
initial field are found to be centered Gaussian random numbers with
variance\[
\left\langle \psi _{k}\psi _{n}\right\rangle =0,\quad \frac{1}{2}D\left(\omega _{k}\right)\left\langle \psi _{k}^{*}\psi _{n}\right\rangle =\delta _{nk}\varepsilon \left(\omega _{k}\right).\]
The amplitudes $\psi _{k}$ relate to the linear energy (\ref{eq:E0}),
and give the smooth function $\varepsilon \left(\omega _{k}\right)=D\left(\omega _{k}\right)\left\langle E_{k}^{\left(0\right)}\bigm /\omega _{k}^{2}\right\rangle $
an interpretation as a frequency-normalized spectral energy density.
The linear amplitudes $a_{k}$ (\ref{eq:a}) form a centered Gaussian
ensemble as well. Their pairwise time correlation is calculated as\begin{equation}
D\left(\omega _{k}\right)\left\langle a_{k}\left(t\right)a_{n}\left(\tau \right)\right\rangle =\delta _{kn}\varepsilon \left(\omega _{k}\right)\cos \omega _{k}\left(t-\tau \right).\label{eq:aStatistics}\end{equation}

Both, mean density of states and correlation matrix, can be calculated
in terms of the average Green's function: $D\left(\omega \right)=2\omega \Im \left(Tr\left\langle \mathbf{G}\left(\omega \right)\right\rangle \right)/\pi $,
and $\mathbf{K}\left(\omega \right)=2\omega \Im \left\langle \mathbf{G}\left(\omega \right)\right\rangle /\pi D\left(\omega \right)$
\cite{foot:GreensDyadic}. For time scales under consideration, in
particular $t_{l}\ll t_{H}$, the characteristic wavelength on the
frequencies of interest is much smaller than the diameter of the solid:
$\lambda /l\ll 1$. The leading order nonlinear contribution thus
comes from the bulk rather than near-boundary region of the solid.
This allows us to neglect effects of the latter, and approximate the
exact Green's function $\mathbf{G}$ in the solid by the Green's function
in an unbounded medium $\mathbf{G}^{\infty }$. This approximation
implies that $\mathbf{K}$ has an infinite correlation radius inherited
from $\mathbf{G}^{\infty }$, and leads to formal integral divergence
in the calculations of the following sections. To mend the problem
we consider the scattering of the field inside the solid as a diffuse
process with a free mean path on the order of the diameter $l$, which
now provides a finite correlation radius for the model: $\left\langle \mathbf{G}\right\rangle =\left\langle \mathbf{G}^{\infty }\right\rangle e^{-\left|\mathbf{x}-\mathbf{x}'\right|/l}$
\cite{ref:Prigodin}. The ansatz is justified as the final results
turn out not to depend on the specific choice of the value of $l$,
as long as it stays much greater than the wavelength.

\section{Energy spectrum evolution}

We assume nonlinear effects to be small and seek solution of equation
(\ref{eq:ODE}) by utilizing small perturbation theory, and expanding
amplitudes $d_{k}$ in orders of magnitude $d_{k}=a_{k}+b_{k}+\left(\ldots \right)$,
where $a_{k}$ are given by (\ref{eq:a}). Next order amplitudes $b_{k}$
arise from the presence of nonlinearity, and are determined solely
by the linear field:\begin{align*}
b_{k}\left(t\right)=-\int _{t_{0}}^{t} & d\tau \frac{\sin \omega _{k}\left(t-\tau \right)}{\omega _{k}}\\
\times \biggl [ & \sum _{m,l}N_{\alpha \beta \gamma }U_{\alpha \beta \gamma }^{klm}\; a_{l}\left(\tau \right)a_{m}\left(\tau \right)\\
+ & \sum _{m,l,n}N_{\alpha \beta \gamma \delta }U_{\alpha \beta \gamma \delta }^{klmn}\; a_{l}\left(\tau \right)a_{m}\left(\tau \right)a_{n}\left(\tau \right)\biggr ].
\end{align*}

To obtain successive corrections to the average spectral density of
the power flow we expand (\ref{eq:PiModal}) in a series of magnitudes
of $a_{k}$:\begin{equation}
\Pi \left(\omega _{k},t\right)\equiv D\left(\omega _{k}\right)\left\langle \Pi _{k}\left(t\right)\right\rangle =\sum _{n}\Pi ^{\left(n\right)}\left(\omega _{k},t\right).\label{eq:PiExpansion}\end{equation}
As mentioned in Section II, since the power flow arises from the nonlinear
mode coupling, the modal energy of the linear field (\ref{eq:E0})
is conserved: $\Pi ^{\left(0\right)}\left(\omega ,t\right)=0$. For
the assumed statistics of the amplitudes $a_{k}$ (\ref{eq:aStatistics})
the first-order correction to the average power density is zero as
well:\begin{align*}
\Pi ^{\left(1\right)} & \left(\omega _{k},t\right)\\
= & D\left(\omega _{k}\right)\sum _{l,m}N_{\alpha \beta \gamma }\left\langle U_{\alpha \beta \gamma }^{klm}\right\rangle \left\langle \dot{a}_{k}\left(t\right)a_{l}\left(t\right)a_{m}\left(t\right)\right\rangle =0.
\end{align*}

The power flow expansion (\ref{eq:PiExpansion}) starts with $\Pi ^{\left(2\right)}$
as the leading term. We express the fourth moments of the amplitudes
$a_{k}$ as double products of their pairwise correlations (\ref{eq:aStatistics}),
and obtain the power flow in terms of the energy density: \begin{widetext}\begin{align}
\Pi ^{\left(2\right)}\left(\omega _{k},t\right)=D\left(\omega _{k}\right)\Biggl \{ & \sum _{l,m}N_{\alpha \beta \gamma }N_{\nu \mu \eta }\left\langle U_{\alpha \beta \gamma }^{kll}U_{\nu \mu \eta }^{kmm}\right\rangle \varepsilon '\left(\omega _{m}\right)\varepsilon '\left(\omega _{m}\right)\frac{\sin \omega _{k}t}{\omega _{k}}\nonumber \\
 & +\frac{1}{2}\sum _{l,m,\pm }N_{\alpha \beta \gamma }N_{\nu \mu \eta }\left\langle U_{\alpha \beta \gamma }^{klm}U_{\nu \mu \eta }^{klm}\right\rangle \varepsilon '\left(\omega _{l}\right)\left[\varepsilon '\left(\omega _{m}\right)\pm 2\varepsilon '\left(\omega _{k}\right)\frac{\omega _{k}}{\omega _{m}}\right]\nonumber \\
 & \qquad \qquad \times \left[\frac{\sin \left(\omega _{k}-\omega _{l}\pm \omega _{m}\right)t}{\omega _{k}-\omega _{l}\pm \omega _{m}}+\frac{\sin \left(\omega _{k}+\omega _{l}\pm \omega _{m}\right)t}{\omega _{k}+\omega _{l}\pm \omega _{m}}\right]\Biggr \},\label{eq:Pi2Integrated}
\end{align}
\end{widetext} where $\varepsilon '\left(\omega \right)=\varepsilon \left(\omega \right)/D\left(\omega \right)$,
is the average normalized energy of a single mode. The modal sum is
evaluated as a frequency integral with integrand weighed by the mean
density of states:\[
\sum _{n}f_{n}=\int _{0}^{+\infty }f\left(\omega _{n}\right)D\left(\omega _{n}\right)d\omega _{n}\]
 We note that matrix $N_{\alpha \beta \gamma \delta }$ responsible
for the cubic nonlinearity does not enter (\ref{eq:Pi2Integrated}),
for its contribution is proportional to $\left\langle \dot{a}_{k}\left(t\right)a_{l}\left(t\right)a_{m}\left(t\right)a_{n}\left(t\right)\right\rangle =0$. 

The centered Gaussian statistics of the modes allows averages in the
form of $\left\langle U_{\alpha \beta \gamma }^{klm}U_{\nu \mu \eta }^{pqr}\right\rangle $
to be expressed as triple products of the pairwise correlations (\ref{eq:UPrimeStatistics}).
Combining terms with the same modal indices we formally write (with
no implied summation on Latin indices) \begin{align}
 & N_{\alpha \beta \gamma }N_{\nu \mu \eta }\left\langle U_{\alpha \beta \gamma }^{klm}U_{\nu \mu \eta }^{klm}\right\rangle \nonumber \\
 & =\mathbb{N}_{0}\left(\omega _{k},\omega _{m},\omega _{l}\right)+\mathbb{N}_{2}\left(\omega _{k},\omega _{l}\right)\delta _{km}+\mathbb{N}_{2}\left(\omega _{m},\omega _{k}\right)\delta _{lm}\nonumber \\
 & \quad +\mathbb{N}_{2}\left(\omega _{l},\omega _{m}\right)\delta _{lk}+\mathbb{N}_{3}\left(\omega _{k}\right)\delta _{kl}\delta _{km}.\label{eq:NFactors}
\end{align}
 The coupling functions $\mathbb{N}$ are defined later.

Expression (\ref{eq:Pi2Integrated}) with the factors (\ref{eq:NFactors})
inserted, though bulky and cumbersome to analyze, gives the leading
term in the power flow due to a weak nonlinearity. In experimental
practice, however, it is not uncommon to deal with the fields that
have the spectral energy density $\varepsilon $ varying smoothly
on frequency scales $\Delta \omega $ greater than the averaging bandwidth
of the limited observation time: $\Delta \omega t\gg 1$. In this
case (\ref{eq:Pi2Integrated}) simplifies further. All the terms directly
proportional to rapidly oscillating sine factors yield negligible
averages:\[
\sin \left(\omega _{n}t\right)/\omega _{n}=D\left(\omega _{n}\right)O\left(1/\Delta \omega t\right),\]
 and the modal sums involving such factors are evaluated as follows:\begin{align*}
\sum _{n}f_{n}\sin \left[\left(\omega _{n}-\omega \right)t\right] & /\left(\omega _{n}-\omega \right)\\
= & \pi D\left(\omega \right)f\left(\omega \right)\left[\theta \left(\omega \right)+O\left(1/\Delta \omega t\right)\right],
\end{align*}
 where $\theta $ is a unit step function. Another simplification
arises if we pay attention only to the frequencies that carried no
energy initially: $\varepsilon \left(\omega \right)=0$. For such
frequencies a weak change in the energy density due to $\Pi ^{\left(2\right)}$
is not masked by a strong initial linear field, and is convenient
for experimental measurement. 

With the simplifications mentioned the power flow expression reduces
to \begin{align}
\Pi ^{\left(2\right)}\left(\omega ,t\right)=\frac{\pi }{2}\sum _{\pm }\int _{0}^{+\infty } & D\left(\omega \right)\mathbb{N}_{0}\left(\omega ,\omega ',\left|\omega \pm \omega '\right|\right)\nonumber \\
 & \times \varepsilon \left(\left|\omega \pm \omega '\right|\right)\varepsilon \left(\omega '\right)d\omega '.\label{eq:Pi20Density}
\end{align}
 The only remaining coupling function is given by contraction of the
nonlinear and correlation matrices: \begin{align}
\mathbb{N}_{0} & \left(\omega _{1},\omega _{2},\omega _{3}\right)\nonumber \\
 & =N_{\alpha \beta \gamma }N_{\nu \mu \eta }\; {K'}_{\alpha \nu }\left(\omega _{1}\right){K'}_{\beta \mu }\left(\omega _{2}\right){K'}_{\gamma \eta }\left(\omega _{3}\right).\label{eq:N0}
\end{align}
 We note that\begin{align*}
\mathbb{N}_{0}\left(\omega ,\omega ',\left|\omega +\omega '\right|\right) & =\mathbb{N}_{0}\left(\omega ,\omega '',\left|\omega -\omega ''\right|\right)\bigr |_{\omega ''=\omega +\omega '}.
\end{align*}

The power flow in the form of (\ref{eq:Pi20Density}) allows simple
interpretation: the energy transfered into a given frequency $\omega $
comes from all pairs of frequencies $\omega '$ and $\omega ''$ that
have the given frequency as a combination, i.e. equal to their sum
or difference: $\omega =\left|\omega '\pm \omega ''\right|$. The
qualitative result is in agreement with and could have been expected
from an elementary theory of nonlinear oscillations \cite{ref:MierovitchBook}.
The symmetry of the nonlinear matrix (\ref{eq:NSymmetry}) leads to
the symmetry of $\mathbb{N}_{0}$ with respect to any interchange
of its arguments. The coupling strength of any triad of frequencies
is thus independent of the energy transfer direction. Nevertheless,
the power itself exhibits an overall trend of the energy to be transfered
up the frequency spectrum, as it is proportional to the density of
states at the target frequency $D\left(\omega \right)$.

The expression (\ref{eq:Pi20Density}) implies energy growth that
is, if $\varepsilon $'s are approximately constant in time, proportional
to elapsed time $t$. In discrete spectrum systems such behavior is
found when a triad of frequencies is locked in internal resonance,
producing secular terms in the solution obtained by regular perturbation
theory. This is the behavior observed in our case because of the finite
time $t\ll t_{H}$: since individual modes are not resolved, any combination
frequency produced by the source frequencies in the frequency range
of interest is indistinguishable from at least one of the normal frequencies
of the system, which hence lies in effective internal resonance with
them.

\section{Estimates for isotropic homogeneous elastic solid}

We return to elastodynamic displacement fields, and specialize to
the case of an elastic body composed of known isotropic homogeneous
material, the case that holds premium experimental and theoretical
interest.

The linear part of the Green's function in the unbounded medium yields
the correlation matrix $\mathbf{K}$ (see Appendix A): \begin{align}
K_{ij} & \left(\Delta \mathbf{x},\omega \right)=\frac{1}{M}e^{-\left|\Delta \mathbf{x}\right|/l}\frac{1}{1/c_{l}^{3}+2/c_{t}^{3}}\nonumber \\
\times  & \biggl \{\frac{\delta _{ij}}{3}\left[\frac{1}{c_{l}^{3}}j_{0}\left(k_{l}\left|\Delta \mathbf{x}\right|\right)+\frac{2}{c_{t}^{3}}j_{0}\left(k_{t}\left|\Delta \mathbf{x}\right|\right)\right]\nonumber \\
 & \quad -\left(\delta _{ij}/3-\Delta \hat{x}_{i}\Delta \hat{x}_{j}\right)\nonumber \\
 & \qquad \times \left[\frac{1}{c_{l}^{3}}j_{2}\left(k_{l}\left|\Delta \mathbf{x}\right|\right)-\frac{1}{c_{t}^{3}}j_{2}\left(k_{t}\left|\Delta \mathbf{x}\right|\right)\right]\biggr \},\label{eq:K}
\end{align}
 where $\Delta \mathbf{x}=\mathbf{x}-\mathbf{x}'$ is a separation
variable, $\Delta \hat{\mathbf{x}}=\Delta \mathbf{x}/\left|\Delta \mathbf{x}\right|$,
$M$ is total mass of the solid, $c_{l}$ and $c_{t}$ are the longitudinal
and transverse wavespeeds respectively, and $j_{n}$ is the spherical
Bessel function of order $n$. The mean density of states for clamped
boundary conditions is calculated in \cite{ref:Dupuis}: \begin{align*}
D\left(\omega \right)= & \frac{V_{0}}{2\pi ^{2}}\omega ^{2}\left[1/c_{l}^{3}+2/c_{t}^{3}\right]\\
 & -\frac{S_{0}}{8\pi c_{l}^{2}}\omega \left[2+\left(c_{l}/c_{t}\right)^{2}+3\left(c_{l}/c_{t}\right)^{4}\right]\\
 & \qquad \Bigm /\left[\left(c_{l}/c_{t}\right)^{2}+1\right]+O\left(l/c\right),
\end{align*}
with $S_{0}$ being the surface area of the solid. 

For the model of nonlinearity at hand appropriate for an isotropic
solid, described by the five-constant theory \cite{ref:MurnaghanBook},
and correlation matrix $\mathbf{K}$ provided by (\ref{eq:K}), the
coupling function $\mathbb{N}_{0}$ (\ref{eq:N0}) assumes the form
\begin{align*}
 & \mathbb{N}_{0}\left(\omega ,\omega ',\left|\omega -\omega '\right|\right)\\
 & =\frac{\pi }{\left(\gamma _{l}^{3}+2\gamma _{t}^{3}\right)^{3}}\frac{c}{V_{0}M}\, \omega ^{3}\, \left|1-\frac{\omega '}{\omega }\right|\frac{\omega '}{\omega }\, \widetilde{\mathbb{N}}\left(\omega '/\omega \right).
\end{align*}
$\widetilde{\mathbb{N}}$ is a dimensionless function of the frequency
ratio and material properties that characterize nonlinear coupling
strength between the source and target frequencies. The function is
independent of the linear dimensions of the body and possesses the
symmetries according to (\ref{eq:NSymmetry}): $\widetilde{\mathbb{N}}\left(\omega '/\omega \right)=\widetilde{\mathbb{N}}\left(\omega /\omega '\right)=\widetilde{\mathbb{N}}\left(1-\omega '/\omega \right)$.
An analytical expression for $\widetilde{\mathbb{N}}$ is available,
but too bulky to be presented here. Details of the calculations are
found in Appendix A.

As a sample distribution of the (linear) energy density we take two
Gaussian peaks with half-width $\Delta \omega $ and total energies
$E_{1,2}$ centered at frequencies $\omega _{1}$ and $\omega _{2}$:\begin{equation}
\varepsilon \left(\omega \right)=\sum _{i=\left\{ 1,2\right\} }\frac{E_{i}}{\omega _{i}^{2}\sqrt{2\pi \Delta \omega ^{2}}}e^{-\left(\omega -\omega _{i}\right)^{2}/2\Delta \omega ^{2}}.\label{eq:epsGauss}\end{equation}
 The width of the peaks is restricted by condition $\Delta \omega t\gg 1$,
as imposed by applicability requirements for the simplified power
input expression (\ref{eq:Pi20Density}). It is additionally assumed
that the peaks do not overlap: $\left|\omega _{1}-\omega _{2}\right|\gg \Delta \omega $.
The suggested form of the spectrum closely models the combined spectra
of two narrow-band signals that might be used in an experiment. By
adjusting the carrying frequencies the features of nonlinear mode
coupling can be investigated, and the form of $\widetilde{\mathbb{N}}$
measured.

As a parameter suitable for characterization of the nonlinear energy
transfer strength we consider transfer times defined as the formal
time required for entire mean energy of the source peaks (\ref{eq:epsGauss})
to be transfered into combination frequencies, provided small perturbation
theory and equation (\ref{eq:Pi20Density}) holds: $t_{1,2}=E_{1,2}\bigm /\Pi _{1,2}^{\left(2\right)}$,
and $t_{1\pm 2}=\sqrt{E_{1}E_{2}}\bigm /\Pi _{1\pm 2}^{\left(2\right)}$
. Here $\Pi ^{\left(2\right)}$ represents the total power input into
frequency band supporting the resulting peak at the combination frequency.
According to the definition $t_{1,2}$, $t_{1+2}$, and $t_{1-2}$
characterize energy transfer into double, sum and difference frequency
respectively. Calculations yield these times to be\begin{align}
t_{1\pm 2}= & \frac{1}{\pi }\frac{\omega _{1}^{2}\omega _{2}^{2}}{Mc^{2}\, D\left(\left|\omega _{1}\pm \omega _{2}\right|\right)\mathbb{N}_{0}\left(\left|\omega _{1}\pm \omega _{2}\right|,\omega _{1},\omega _{2}\right)}\frac{1}{m_{1\pm 2}^{2}},\nonumber \\
 & \label{eg:transferTs}\\
t_{1,2}= & \frac{2}{\pi }\frac{\omega _{1,2}^{4}}{Mc^{2}\, D\left(2\omega _{1,2}\right)\mathbb{N}_{0}\left(2\omega _{1,2},\omega _{1,2},\omega _{1,2}\right)}\frac{1}{m_{1,2}^{2}}.\nonumber 
\end{align}
 As a measure of absolute strength of the linear field we choose squares
of the Mach numbers: $m_{1\pm 2}^{2}=\sqrt{E_{1}E_{2}}/Mc^{2}$ and
$m_{1,2}^{2}=E_{1,2}/Mc^{2}$. As seen from (\ref{eg:transferTs})
stronger nonlinear effects (shorter times) require higher Mach numbers,
i.e. stronger linear fields.

\begin{figure}
\includegraphics{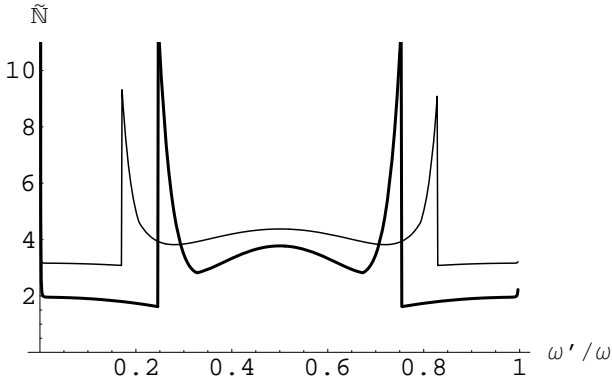}

\caption{\label{fig:N} Dimensionless coupling function in aluminum (thick
line) and fused silica (thin line).}
\end{figure}

For numerical estimates we take an aluminum block and a fused silica
block with the mechanical properties listed in Table \ref{table:Properties},%
\begin{table*}

\caption{\label{table:Properties}Mechanical properties.}

\begin{tabular}{ccccccccc}
\hline 
\hline&
$\rho _{0}\left(\text {kg}/\text {m}^{3}\right)$&
$c_{l}\left(\text {m}/\text {s}\right)$&
$c_{t}\left(\text {m}/\text {s}\right)$&
$\lambda \left(\text {GPa}\right)$&
$\mu \left(\text {GPa}\right)$&
$A\left(\text {GPa}\right)$&
$B\left(\text {GPa}\right)$&
$C\left(\text {GPa}\right)$\\
\hline 
Aluminum\footnote[1]{Alloy D54S, Smith \emph{et al.} \cite{ref:Smith}}&
$2720$&
$6100$&
$3090$&
$49.1$&
$26.0$&
$-320$&
$-198$&
$-190$\\
Fused Silica&
$2200$\footnote[2]{Drumheller \cite{ref:Drumheller}}&
$5700$&
$3750$&
$9.60$\footnotemark[2]&
$30.9$\footnotemark[2]&
$-44$\footnote[3]{Bechmann \emph{et al.} \cite{ref:Bechmann}}&
$93$\footnotemark[3]&
$27$\footnotemark[3]\\
\hline
\end{tabular}
\end{table*}
 and choose the transverse wavespeed for each material as the characteristic
one: $c=c_{t}$. Due to the inversion symmetry we evaluate the coupling
function $\widetilde{\mathbb{N}}$ for frequency range $0<\omega '/\omega <1$
only, and plot it in Figure \ref{fig:N}.%
 Immediately a significant feature of the plot comes into view: the
sharp peaks and discontinuities of the coupling function at frequency
ratios of $0.25$ and $0.75$ for aluminum, and $0.17$ and $0.83$
for fused silica.

The feature may be understood by considering nonlinear combinations
of plane waves. We recall that from a plane wave perspective appropriate
in the limit $\lambda /l\ll 1$, in addition to internal frequency
resonance between the source frequencies $\omega '$ and $\omega ''=\left|\omega -\omega '\right|$
and the target frequency $\omega $, there is an additional requirement
for wavevector resonance. We note that there are two plane wave types
in the system, namely longitudinal and transverse. If the types of
the three waves (at $\omega $, $\omega '$ and $\omega ''$) are
identical, wavevector resonance ($\mathbf{k}\left(\omega ''\right)=\mathbf{k}\left(\omega \right)+\mathbf{k}\left(\omega '\right)$
with $\left|\mathbf{k}\right|=\omega /c$) is always possible, and
demands that the three wavevectors are parallel or antiparallel. If
the wave type for one of the frequencies is different from the other
two, then wavevector resonance is not always possible; it depends
on the frequency ratio. If it is possible, there will be a nontrivial
angle between the wavevectors. The transition between possibility
and impossibility occurs at certain special frequency ratios, $\omega /\omega '=\left(1\pm c_{t}/c_{l}\right)/2$.
At these ratios wavenumber resonance occurs with parallel or antiparallel
wavevectors. The special frequency ratio depends solely on the Poisson
ratio of the material. For non-dispersive single wavespeed systems,
when $c_{l}=c_{t}$, the ratios become $0$ and $1$. In this case
the effect is not observed, as one of the source frequencies needs
to be zero \cite{footScalar}. However, for certain dispersive single
wavespeed systems the peaks and discontinuities might still be found,
if the dispersion equation of the system is such that the wavenumber
resonance is possible. It is also worth noting that the discontinuity
position at leading order is independent of the field strength. Thus
field calibration is not required in experimental measurements in
order to observe this characteristic feature, for only relative values
of $\widetilde{\mathbb{N}}$ are needed. This makes the method convenient
to use in the light of its eventual application to NDE.

\begin{figure}
\includegraphics{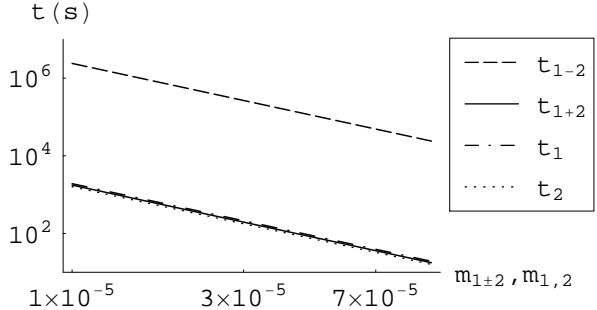}

\caption{\label{fig:T} Transfer times in aluminum block for fixed $\omega _{1}/2\pi =500\, \text {kHz}$
and $\omega _{2}/2\pi =600\, \text {kHz}$.}
\end{figure}

In order to see whether the presented theory is applicable for experimental
verification and eventual use, we need to provide numerical estimates
for transfer times (\ref{eg:transferTs}). For this purpose we choose
the solid to be equivalent in volume and surface area to a cube with
a side of $7\text {cm}$. The typical values of the transition (ballistic)
time in the system are of the order $10\, \mu s$. The carrying frequencies
of the two narrow-band signals are taken to be $\omega _{1}/2\pi =500\, \text {kHz}$
and $\omega _{2}/2\pi =600\, \text {kHz}$, so that the break (Heisenberg)
time is of the same order as the absorption times of $100\, \text {ms}$
common to experiment. The transfer times are calculated for the aluminum
block and plotted in Figure \ref{fig:T} as functions of the linear
field strength, characterized by Mach number.%

It is feasible to generate linear diffuse fields with RMS elastic
strains corresponding to Mach numbers of order $10^{-5}$ that yield
transfer times of order $10^{3}\, \text {s}$ (see Figure \ref{fig:T}).
Thus the observation times of the order of the absorption time yield
energy densities at combination frequencies some $10^{4}$ times weaker
than the source signal energy, i.e. RMS strains of $10^{-7}$. These
are easily detectable. We also note that the strain ratio (combination
field to initial field) is much smaller than unity, so the use of
regular perturbation theory is valid for the given time scale.

\begin{figure}
\includegraphics{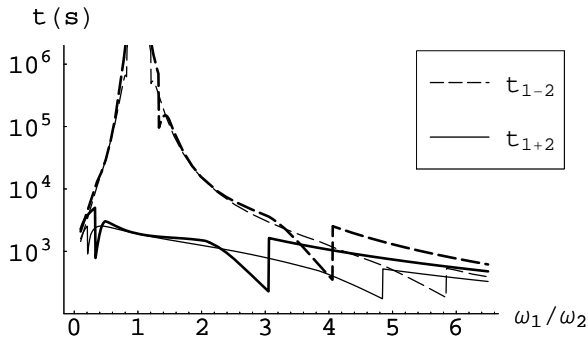}

\caption{\label{fig:TFixed} Transfer times in aluminum (thick lines) and
fused silica (thin lines) block for varying source frequency ratio,
and fixed $\omega _{2}/2\pi =550\, \text {kHz}$ and $m_{1\pm 2}=10^{-5}$.}
\end{figure}

Finally, we calculate transfer times in aluminum and fused silica
as a function of source frequency ratio for a fixed Mach number (see
Figure \ref{fig:TFixed}). %
 The frequency dependence of the nonlinear coupling strength in terms
of the transfer times reveals the frequency dependence of $\widetilde{\mathbb{N}}$.
The dips in the transfer times correspond to the peaks of the coupling
function, and might be used to estimate Poisson's ratio of the material,
or as a signature of nonlinearity. We notice that the energy transfer
into near-zero frequencies is inefficient, as manifested by high difference
frequency transfer times in the vicinity of $\omega _{1}/\omega _{2}=1$.

\section{Conclusion}

In the present work we have obtained a formula for the evolution of
the average linear energy spectrum (\ref{eq:Pi20Density}) of a weakly
nonlinear system. Except for definition of the modal linear energy
$E_{k}$ (\ref{eq:E}), and statistical properties of the normal frequencies
$\omega _{k}$ and modes $\mathbf{u}^{k}$ (\ref{eq:EigenProblem}),
the result has no reference to the physical nature of the dynamic
system governed by nonlinear equations (\ref{eq:ODE}). With proper
generalization of the matrices $\mathbf{NU}$ describing the physical
nature of the nonlinearity it might be applicable beyond the scope
of elasticity. 

We observe that the energy redistribution occurs for triads of frequencies,
with one being the sum or difference of the other two. The average
power input into a narrow frequency band is found to be a cumulative
effect coming from such interactions in the initial energy spectrum
of the field, and to be proportional to the convolution of the energies
stored at two frequencies. Relative weight of the interactions is
given by the frequency-dependent coupling function $\mathbb{N}_{0}$
(\ref{eq:N0}), and is calculated as a contraction of the nonlinear
and correlation matrices $\mathbf{N}$ and $\mathbf{K}'$. 

The case of chief experimental interest involving an isotropic homogeneous
elastic body with nonlinearity given by the five-constant theory and
an initial energy spectrum consisting of two narrow-band signals is
discussed in detail. We find that characteristic times for the full
energy transfer from the source into combination frequencies depend
on the ratio of the source frequencies and exhibit characteristic
dips and peaks and discontinuities at special frequency ratios corresponding
to wavenumber resonance. The position of the resonance depends on
Poisson's ratio. We also find that the transfer times are reciprocal
to the strength of the initial signal given by the square of the corresponding
Mach number. 

The current theory is derived for zero-displacement boundary conditions
of a clamped solid, and is impractical for experimental realization.
However, for the frequencies of interest the main contribution to
the mixing comes from the bulk of the solid. Since the near-boundary
regions play a lesser role, the authors believe that the energy spectrum
evolution for the traction-free elastic solid would exhibit behavior
similar to that discussed here, and thus be accessible for experimental
verification.

\appendix

\section{Correlation matrix and coupling function for isotropic homogeneous
medium}

Elastic deformation of an isotropic homogeneous solid is described
by the five-constant theory in terms of the Lam\'{e} constants $\lambda $
and $\mu $, and nonlinear coefficients $A$, $B$ and $C$ \cite{ref:MurnaghanBook}.
According to the theory the linear and second-order nonlinear elastic
tensors have the form of\begin{align}
C_{ijkl}= & \lambda \delta _{ij}\delta _{kl}+2\mu \Phi _{ijkl}^{1},\nonumber \\
 & \label{eq:ElasticTensors}\\
D_{ijklmn}= & 2A\Phi _{ijklmn}^{2}+2C\delta _{ij}\delta _{kl}\delta _{mn}\nonumber \\
 & +2B\left(\delta _{ij}\Phi _{klmn}^{1}+\delta _{kl}\Phi _{ijmn}^{1}+\delta _{mn}\Phi _{ijkl}^{1}\right).\nonumber 
\end{align}
 The elementary isotropic tensors $\Phi $ are given as follows:\[
\Phi _{ijkl}^{1}=\frac{1}{2}\bigl (\delta _{ik}\delta _{jl}+\delta _{il}\delta _{jk}\bigr ),\]
\begin{align*}
\Phi  & _{ijklmn}^{2}\\
 & =\frac{1}{8}\bigl (\delta _{ik}\delta _{jm}\delta _{ln}+\delta _{ik}\delta _{jn}\delta _{lm}+\delta _{il}\delta _{jm}\delta _{kn}+\delta _{il}\delta _{jn}\delta _{km}\\
 & \qquad +\delta _{im}\delta _{jk}\delta _{ln}+\delta _{im}\delta _{jl}\delta _{kn}+\delta _{in}\delta _{jk}\delta _{lm}+\delta _{in}\delta _{jl}\delta _{km}\bigr ).
\end{align*}
 By substituting tensors (\ref{eq:ElasticTensors}) into the strain
energy definition (\ref{eq:EnergyDensity}), particular form of directional
tensors $N_{ijklmn}$ (\ref{eq:NTensor}) is derived.

The Green's function in the unbounded medium $\mathbf{G}^{\infty }$
is calculated from its spatial Fourier-transform \cite{ref:Weaver90}:
\[
G_{ij}^{\infty }\left(\mathbf{p},\omega \right)=\frac{\hat{p}_{i}\hat{p}_{j}}{c_{l}^{2}p^{2}-\omega ^{2}}+\frac{\delta _{ij}-\hat{p}_{i}\hat{p}_{j}}{c_{t}^{2}p^{2}-\omega ^{2}},\quad \hat{\mathbf{p}}=\mathbf{p}/\left|\mathbf{p}\right|.\]
Direct integration of the above expression and its subsequent normalization
(\ref{eq:KNorm}) yields particular expression for correlation matrix
$\mathbf{K}$ (\ref{eq:K}). As two of its particular limit cases
we note, first, the known autocorrelation function for scalar Helmholtz
equation: $j_{0}\left(k\left|\Delta \mathbf{x}\right|\right)$ \cite{ref:Stockmann}
obtained by letting $c=c_{l}=c_{t}$. And second, the autocorrelation
function for purely transverse field \---  such as, for example,
electromagnetic field \cite{ref:Eckhardt} \---  obtained by letting
$c_{l}\rightarrow \infty $.

The correlation matrix of the first partial derivatives of the modes
is derived from $\mathbf{K}$ by means of (\ref{eq:KPrimeCalc}).
It is expressed in terms of known directional tensors $\mathbf{H}$,
characteristic wavenumber $k=\omega /c$, and wavespeed ratios $\gamma _{l,t}=c/c_{l,t}$:
\begin{align}
K'{}_{\alpha \equiv \left\{ \mathbf{x},i,l\right\} \beta \equiv \left\{ \mathbf{x}',j,m\right\} } & =\frac{1}{M}e^{-\left|\Delta \mathbf{x}\right|/l}\nonumber \\
\times \frac{1}{\gamma _{l}^{3}+2\gamma _{t}^{3}} & k^{2}\sum _{{{a=\left\{ 0,2,4\right\} \atop \mathfrak{p}=\left\{ l,t\right\} }}}\gamma _{\mathfrak{p}}^{5}H_{ijmn}^{\left(\mathfrak{p},a\right)}\, j_{a}\left(\gamma _{\mathfrak{p}}k\left|\Delta \mathbf{x}\right|\right).\label{eq:KPrime}
\end{align}
 The sum over longitudinal and transverse wave types is denoted as
$\mathfrak{p}=\left\{ l,t\right\} $. Directional tensors $\mathbf{H}$
are defined by the following expressions:\begin{align*}
H_{ijmn}^{\left(l,0\right)} & =\frac{1}{3}\delta _{ij}\delta _{mn}-H_{ijmn}^{\left(t,0\right)}\\
 & =\frac{1}{15}\left[\delta _{ij}\delta _{mn}+\delta _{im}\delta _{jn}+\delta _{in}\delta _{jm}\right],
\end{align*}
\[
H_{ijmn}^{\left(l,2\right)}=\delta _{ij}Q_{mn}^{2}-H_{ijmn}^{\left(t,2\right)}=\frac{1}{7}\, Q_{ijmn}^{2},\]
\[
H_{ijmn}^{\left(l,4\right)}=-H_{ijmn}^{\left(t,4\right)}=\frac{1}{7}Q_{ijmn}^{2}-Q_{ijmn}^{4}.\]
 The quadruple, composite quadruple and $2^{4}$-order directional
moments \textbf{$\mathbf{Q}$} moments yield zero values when integrated
upon all spatial directions or contracted upon any two pairs of indices:
\begin{align*}
Q_{ij}^{2}= & \frac{\delta _{ij}}{3}-\Delta \hat{x}_{i}\Delta \hat{x}_{j},\\
Q_{ijmn}^{2}= & \delta _{ij}Q_{mn}^{2}+\delta _{mn}Q_{ij}^{2}+\delta _{im}Q_{jn}^{2}+\delta _{jn}Q_{im}^{2}\\
 & +\delta _{in}Q_{jm}^{2}+\delta _{jm}Q_{in}^{2},\\
Q_{ijmn}^{4}= & \frac{1}{15}\left(\delta _{ij}\delta _{mn}+\delta _{im}\delta _{jn}+\delta _{in}\delta _{jm}\right)\\
 & -\Delta \hat{x}_{i}\Delta \hat{x}_{j}\Delta \hat{x}_{m}\Delta \hat{x}_{n}.
\end{align*}

The coupling function (\ref{eq:N0}) is given by contraction of the
directional tensors of the nonlinear matrix (\ref{eq:NTensor}) with
correlation matrices of the first derivatives (\ref{eq:KPrime}).
It is used to find the dimensionless function \begin{align*}
\widetilde{\mathbb{N}}= & \sum _{{{a,b,c=\left\{ 0,2,4\right\} \atop \mathfrak{p},\mathfrak{q},\mathfrak{r}=\left\{ l,t\right\} }}}\left(\gamma _{\mathfrak{p}}\gamma _{\mathfrak{q}}\gamma _{\mathfrak{r}}\right)^{4}\mathbb{H}_{abc}^{\mathfrak{pqr}}I_{abc}^{\mathfrak{pqr}}\left(\omega '/\omega \right),
\end{align*}
where constants $\mathbb{H}$ are \begin{align*}
\mathbb{H}_{abc}^{\mathfrak{pqr}} & =N_{ijklmn}N_{pqrstu}H_{ipjq}^{\left(\mathfrak{p},a\right)}H_{krls}^{\left(\mathfrak{q},b\right)}H_{mtnu}^{\left(\mathfrak{r},c\right)},
\end{align*}
 and the integral $I$ is given by \begin{align}
{I}_{abc}^{\mathfrak{pqr}} & \left(\Omega \right)\nonumber \\
= & \frac{\gamma _{\mathfrak{p}}\gamma _{\mathfrak{q}}}{\gamma _{\mathfrak{r}}^{2}}\, \Omega \left|1-\Omega \right|\int _{0}^{+\infty }z^{2}e^{-z/3\gamma _{\mathfrak{r}}k}\nonumber \\
 & \times j_{a}\left(\frac{\gamma _{\mathfrak{p}}}{\gamma _{\mathfrak{r}}}\Omega z\right)j_{b}\left(\frac{\gamma _{\mathfrak{q}}}{\gamma _{\mathfrak{r}}}\left|1-\Omega \right|z\right)j_{c}\left(z\right)dz.\label{eq:I}
\end{align}

For a finite correlation radius of the coupling matrix $l$, the integral
additionally depends on the target frequency $\omega $. The dependence
turns out to be only significant in the small vicinity of $\Omega =\omega '/\omega =\left\{ 0,1\right\} $,
elsewhere the contribution being small: $O\left(1/kl\right)$. We
note that for the mentioned ratios $\Omega $ one of the source frequencies
$\omega '$ or $\left|\omega -\omega '\right|$ must be close to zero,
and the phenomenon is of small practical importance from experimental
point of view. For these near-zero frequencies the wavelength becomes
comparable to or greater than the diameter of the solid. Thus the
substitution of the exact Green's function $\mathbf{G}$ by the Green's
function in the unbounded medium $\mathbf{G}^{\infty }$ is no longer
valid, and conditions for the time scales made in Section III are
not met. 

With these limitations in mind we set $\lambda /l=0$ for practical
calculations, and obtain analytical expression for (\ref{eq:I}).
As the result of this procedure a singularity of the function $\widetilde{\mathbb{N}}$
at $\omega '/\omega =\left\{ 0,1\right\} $ is acquired (not shown
in Figure \ref{fig:N}). Discontinuities of the function at the wavenumber
resonances, and discontinuities of its slope at $\omega /\omega '=2/\left(1+c_{l}/c_{t}\right)$
and $1-2/\left(1+c_{l}/c_{t}\right)$, are found as well. We expect
in practice to observe sharp transitions over finite ranges in $\omega /\omega '$
of order $1/kl$, at a characteristic frequency ratio $\omega /\omega '$
possibly shifted by an amount $O\left(1/kl\right)$. 

\begin{thebibliography}{10}
\bibitem{ref:Sutin-Zaitsev}V.Yu. Zaitsev, A.M. Sutin, I.Yu. Belyaeva, and V.E. Nazarov, J. Vibration
and Control \textbf{1}(3), 335 (1995); Alexander M. Sutin and Dimitri
M. Donskoy, in \emph{Proceedings of the 8th International Symposium
on Nondestructive Characterization of Materials, Boulder, Colorado,
1997}, edited by Robert E. Green Jr. (Plenum Press, New York, 1998)
\bibitem{ref:Wegner-Rothenfusser}A. Wegner, A. Koka, K. Janser, U. Netzelmann, S. Hirsekorn, and W.
Arnold, Ultrasonics \textbf{38}, 316 (2000); M. Rothenfusser, M. Mayr,
and J. Baumann, \emph{ibid.} \textbf{38}, 322 (2000)
\bibitem{ref:Hurley}D.C. Hurley, J. Acoust. Soc. Am. \textbf{106}(4), 1782 (1999)
\bibitem{ref:deLima}W.J.N. de Lima and M.F. Hamilton, J. Sound Vib. \textbf{265}(4), 819
(2003)
\bibitem{ref:Stockmann}Hans-J\"{u}rgen St\"{o}ckmann, \emph{Quantum Chaos: an Introduction},
(Cambridge University Press, 1999), Chap. 2, 6
\bibitem{ref:Pine-Weaver}D.J. Pine, D.A. Weitz, P.M. Chaikin, and E. Herbolzheimer, Phys. Rev.
Lett. \textbf{60}(12), 1134 (1988); Richard L. Weaver and Wolfgang
Sachse, J. Acoust. Soc. Am. \textbf{97}(4), 2094 (1995)
\bibitem{ref:Boer}Johannes F. deBoer, Ad Lagendijk, Rudolf Sprik, and Shechao Feng,
Phys. Rev. Lett. \textbf{71}(24), 3947 (1993)
\bibitem{ref:Bressoux-Skipetrov}R. Bressoux and R. Maynard, Europhysics Letters \textbf{50}(4), 460
(2000); S.E. Skipetrov and R. Maynard, Phys. Rev. Lett. \textbf{85}(4),
736 (2000); S.E. Skipetrov, Phys. Rev. E \textbf{63}, 056614 (2001)
\bibitem{ref:OgdenBook}R.W. Ogden, \emph{Nonlinear Elastic Deformations} (Ellis Horwood Ltd.,
Chichester, 1984), Chap. 3.4, 4.3, 6.1
\bibitem{foot:BC}Under zero-displacement boundary conditions solution of the nonlinear
problem belongs to the function space with a basis formed by the eigenfunctions
of the linear operator, thus allowing the proposed decomposition.
For zero-traction boundary conditions such decomposition is no longer
valid, since the eigenfunctions no longer span the entire solution
space of the nonlinear problem.
\bibitem{ref:Weaver84}R.L. Weaver, J. Sound Vib. \textbf{94}(3), 319 (1984)
\bibitem{ref:Weaver89}R.L. Weaver, J. Acoust. Soc. Am. \textbf{85}(3), 1005 (1989)
\bibitem{ref:Ellegaard-Schaadt}C. Ellegaard, T. Guhr, K. Lindemann, H.Q. Lorensen, J. Nyg{\aa}rd,
and M. Oxborrow, Phys. Rev. Lett. \textbf{75}(8), 1546 (1995); C.
Ellegaard, T. Guhr, K. Lindemann, J. Nyg{\aa}rd, and M. Oxborrow,
\emph{ibid.} \textbf{77}(24), 4918 (1996); A. Andersen, C. Ellegaard,
A. D. Jackson, and K. Schaadt, Phys. Rev. E \textbf{63}, 066204 (2001)
\bibitem{ref:Berry-OConnor}M.V. Berry, J. Phys A \textbf{10}(12), 2083 (1977); P. O'Connor, J.
Gehlen, and E.J. Heller, Phys. Rev. Lett. \textbf{58}(13), 1296 (1987)
\bibitem{ref:McDonald-Schaadt}Steven W. McDonald and Allan N. Kaufmann, Phys. Rev. A \textbf{37}(8),
3067 (1988); K. Schaadt, T. Guhr, C. Ellegaard, and M. Oxborrow, Phys.
Rev. E \textbf{68}, 036205 (2003)
\bibitem{foot:GreensDyadic}The expressions are obtained by calculating ensemble average of the
imaginary part of the Green's function $\Im G_{ij}\left(\mathbf{x},\mathbf{x}',\omega \right)=\pi \sum _{n=1}^{\infty }u_{i}^{\left(n\right)}\left(\mathbf{x}\right)u_{j}^{\left(n\right)}\left(\mathbf{x}'\right)\delta \left(\omega -\omega _{n}\right)/2\omega $;
E.N Economou, \emph{Green's functions in quantum physics} (Springer-Verlag,
New York, 1979), Chap. 1
\bibitem{ref:Prigodin}V.N. Prigodin, B.L. Altshuler, K.B. Efetov, and S. Iida, Phys. Rev.
Lett. \textbf{72}(4), 546 (1994); V.N. Prigodin, \emph{ibid.} \textbf{74}(9),
1566 (1995)
\bibitem{ref:MierovitchBook}Leonard Mierovitch, \emph{Elements of Vibration Analysis} (McGraw-Hill,
New York, 1986), Chap. 10
\bibitem{ref:Dupuis}Marc Dupuis, Robert Mazo, and Lars Onsager, J. Chem. Phys. \textbf{33}(5),
1452 (1960)
\bibitem{ref:MurnaghanBook}F.D. Murnaghan, \emph{Finite Deformations of an Elastic Solid} (Chapman
\& Hall, New York, 1951), Chap. 4
\bibitem{ref:Smith}R.T Smith, R. Stern, and R.W.B Stephens, J. Acoust. Soc. Am. \textbf{40}(5),
1002 (1966)
\bibitem{ref:Drumheller}D.S. Drumheller, \emph{Introduction to Wave Propagation in Nonlinear
Fluids and Solids} (University Press, Cambridge, 1998), App. B
\bibitem{ref:Bechmann}R. Bechmann and R.F.S. Hearmon, in \emph{Elastic, Piezoelectric, Piezooptic,
Electrooptic Constants, and Nonlinear Dielectric Susceptibilities
of Crystals}, \emph{}edited by K.H. Hellwege and A.M. Hellwege, Landolt-B\"{o}rnstein,
New Series (Springer-Verlag, New York, 1969), Vol. III/2
\bibitem{footScalar}For this reason a discussion of weak nonlinearity in such simple systems
would offer little insight and few rich behaviours.
\bibitem{ref:Weaver90}R.L. Weaver, J. Mech. Phys. Solids \textbf{38}(1), 55 (1990); J.D.
Achenbach, \emph{Wave Propagation in Elastic Solids} (Elsevier, Amsterdam,
1999), Chap. 3
\bibitem{ref:Eckhardt}B. Eckhardt, U. D\"{o}rr, U. Kuhl, and H.-J. St\"{o}ckmann, Europhys.
Lett. \textbf{46}(2), 134 (1999)
\end{thebibliography}
\end{document}